\documentclass[twocolumn]{aastex631}
\usepackage{amsmath}
\usepackage{CJK}
\usepackage{multirow}
\usepackage{soul}

\begin{document}
\begin{CJK*}{UTF8}{gbsn}
\title{
Disk2Planet: A Robust and Automated Machine Learning Tool for Parameter Inference in Disk-Planet Systems
}

\author[0009-0006-0889-3132]{Shunyuan Mao (毛顺元)}
\affiliation{Department of Physics and Astronomy, University of Victoria, Victoria, BC V8P 5C2, Canada, symao@uvic.ca}

\author[0000-0001-9290-7846]{Ruobing Dong (董若冰)}
\affiliation{Department of Physics and Astronomy, University of Victoria, Victoria, BC V8P 5C2, Canada}
\affiliation{Kavli Institute for Astronomy and Astrophysics, Peking University, Beijing 100871, People's Republic of China, rbdong@pku.edu.cn}

\author[0000-0001-9036-3822]{Kwang Moo Yi}
\affiliation{Department of Computer Science, University of British Columbia, Vancouver, BC V6T 1Z4, Canada}

\author[0000-0002-5476-5768]{Lu Lu}
\affiliation{Department of Statistics and Data Science, Yale University, New Haven, CT 06511, USA}

\author{Sifan Wang}
\affiliation{Graduate Group in Applied Mathematics and Computational Science, University of Pennsylvania, Philadelphia, PA 19104, USA}
\affiliation{Institute for Foundations of Data Science, Yale University, New Haven, CT 06520, USA}

\author[0000-0002-2816-3229]{Paris Perdikaris}
\affiliation{Department of Mechanical Engineering and Applied Mechanics, University of Pennsylvania, Philadelphia, PA 19104, USA}

\begin{abstract}

We introduce Disk2Planet, a machine learning-based tool to infer key parameters in disk-planet systems from observed protoplanetary disk structures. Disk2Planet takes as input the disk structures in the form of two-dimensional density and velocity maps, and outputs disk and planet properties, that is, the Shakura--Sunyaev viscosity, the disk aspect ratio, the planet--star mass ratio, and the planet's radius and azimuth. We integrate the Covariance Matrix Adaptation Evolution Strategy (CMA--ES), an evolutionary algorithm tailored for complex optimization problems, and the Protoplanetary Disk Operator Network (PPDONet), a neural network designed to predict solutions of disk--planet interactions. Our tool is fully automated and can retrieve parameters in one system in three minutes on an Nvidia A100 graphics processing unit. We empirically demonstrate that our tool achieves percent-level or higher accuracy, and is able to handle missing data and unknown levels of noise.

\end{abstract}


\section{Introduction} \label{sec: intro}
\end{CJK*}
Planets form in protoplanetary disks: flattened, gaseous disks surrounding newborn stars \citep{andrews2020observations}.  To study planet formation in the outer disk beyond the snowline, a straightforward approach is to directly image the forming planets in these disks \citep[e.g.,][]{keppler2018discovery,muller2018orbital,wagner2018magellan,christiaens2019separating,haffert2019two,hashimoto2020accretion,wang2020keck,currie2022images,zhou2022hst,wagner2023direct}. Successful detections can provide critical information about the process, such as the mass and orbit of the forming planets. However, this approach is challenging with current observational techniques as the signals from the stars and disks overwhelm those from the planets \citep{currie2022images}. These factors severely limit our understanding of how planets form.

To detect young planets embedded in protoplanetary disks, an alternative method is to infer their presence and constrain their parameters from observations of large-scale disk structures produced by disk-planet interactions \citep{paardekooper2022planet}. The morphology of these structures, such as gaps \citep[e.g.,][]{paardekooper2006dust,rosotti2016minimum,dipierro2017opening}, spiral arms \citep[e.g.,][]{bae2018planet,dong2015observational}, vortices \citep[e.g.,][]{zhu2014particle}, and kinematic perturbations \citep[e.g.,][]{pinte2018kinematic,teague2018kinematical,izquierdo2021disc,rabago2021constraining}, depends on the properties of the disk and the planets, such as the viscosity and aspect ratio of the disk, and the masses and orbits of the planets.

Deriving planetary parameters from observed disk structures requires solving {\it the inverse problem}: deducing causes from observed effects. It involves fitting observations with parametrized disk-planet models predicted by a {\it forward problem solver}, a solver that calculates the effects from the causes. The process iteratively fine-tunes parameters to fit the observed data.

\begin{figure}[tb]
    \centering
    \includegraphics[width=0.75\linewidth]{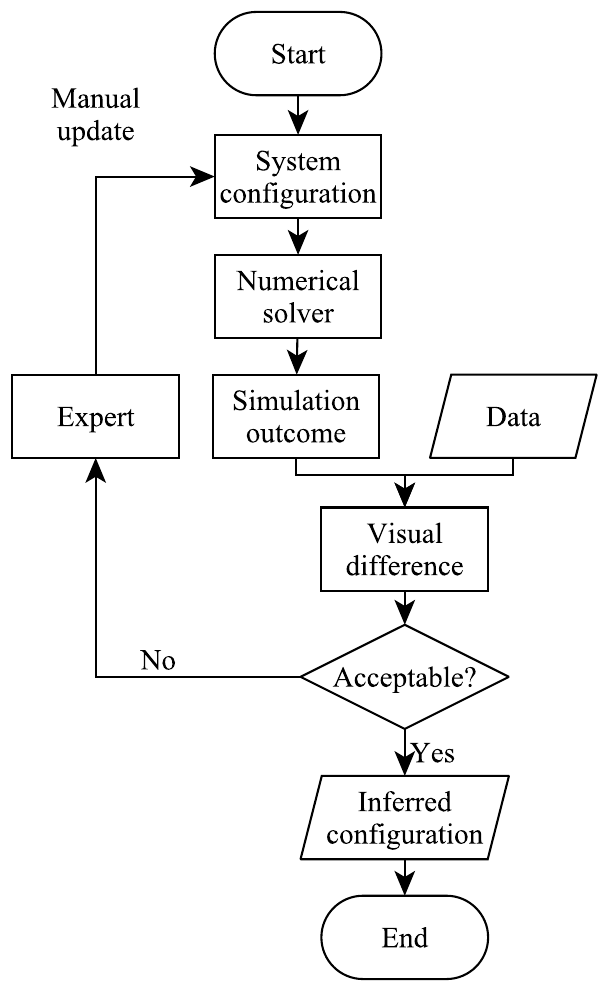}
    \caption{
    The framework of conventional inverse problem solvers. Each iteration begins with an estimation of the system configuration, which includes parameters in a disk-planet system. This configuration is sent to a numerical solver to generate the resulting disk maps. Expert analysis then evaluates the visual discrepancies between the simulation outcomes and the observational data, leading to iterative refinements of the system parameters.
    \label{fig: traditional framework}}
\end{figure}

Conventional inverse problem solvers, as illustrated in Fig. \ref{fig: traditional framework}, are both computationally and labor-wise intensive. The forward problem solver relies on numerical simulations, each demanding substantial computational resources \citep[e.g.,][]{cilibrasi2023meridional}. This demand is multiplied by the number of iterations needed to process a single observed system \citep[e.g.,][]{dipierro2015planet,dong2015observational2,jin2016modeling,pinte2019kinematic}. Additionally, because of the expensive computing, to make the most out of each run, the procedure requires significant expert involvement for the initial setup, continuous monitoring, and adjustments of simulations. As the discovery of structures in disks accelerates---with over $100$ disks newly resolved in a regular year \citep{benisty2022optical}---the demand to fit all disk observations through simulations vastly outstrips the available computational and human resources.

Naturally, alternative approaches that aim to reduce computation has been explored, such as those that directly map planet-induced structures to parameters in disk-planet systems without numerical simulations. Initially, this mapping involved using empirical formulas to analytically correlate a small number of (or a single) scalar observable metrics of disk structures, such as gap width or depth, to system parameters \citep[e.g.,][]{dong2018multiple, zhang2018disk,dong2017bright}. Recently, the same correlations are learnt by neural networks \citep{auddy2020machine,auddy2022using} instead. However, all such approaches rely solely on a limited number of scalar metrics, failing to encapsulate the rich information in images, resulting in degeneracies and large uncertainties in the results (e.g., \citealt{fung2014empty}).

To address these limitations, Convolutional Neural Networks (CNNs) \citep{o2015introduction,li2021survey} are employed in several pioneering works to map entire images to disk and planet parameters \citep{auddy2021dpnnet,zhang2022pgnets,terry2022locating,ruzza2024dbnets}. The CNNs establish the relationships between disk density maps or synthetic observations (``effects'') and system parameters (``causes'') by training on simulated data where both are known. These networks analyze spatial correlations among pixel values to identify the characteristics of planet-induced disk structures, leveraging the comprehensive information in an image rather than relying on a limited array of highly synthesized scalar metrics (e.g., gap width). This method not only improves inference accuracy but also streamlines the process of retrieving planet parameters from disk observations for end users by (1) obviating the need for numerical simulations and (2) circumventing the tedious iterative fitting process.

However, the accuracy of parameters retrieved by CNN-based inverse problem solvers typically only reaches the order of $0.1$. Furthermore, such accuracy is assured solely for images within the training data distribution and may significantly deteriorate when applied to real observations, which may not conform to this distribution. This is particularly evident with images (1) with noise \citep[\S 5.3]{ruzza2024dbnets}, as the noise level in actual observations is measured only afterward and varies from one observation to another, challenging the preparation of the training dataset; and (2) with missing parts, e.g., due to foreground absorption (\citealt{bruderer2014gas}, Fig.~1; \citealt{tsukagoshi2019flared}). These scenarios underscore the inherent difficulties in applying CNN-based solvers to real observations.

Given the traditional methods (e.g., Fig.~\ref{fig: traditional framework}) are computationally and labor-wise expensive, and the CNN-based solvers are limited in accuracy and robustness, a better approach is needed. We address these issues by developing a fast, and fully automated inverse problem solver, named Disk2Planet, that can infer planet and disk parameters from observed disk structures with unprecedented accuracy and robustness. It replaces cumbersome simulations with PPDONet \citep{mao2023ppdonet}, a fast machine learning-based solver capable of predicting one system in approximately $0.01$ seconds on an Nvidia A100 chip.
In addition, Disk2Planet is the first in the field of planet formation to utilize CMA--ES \citep{hansen1996adapting,hansen1997convergence,hansen2001completely,hansen2003reducing,hansen2006cma,hansen2019pycma} for iterative parameter updates, automating the parameter inference process and eliminating the need for human intervention.

Disk2Planet can process one system with a single image input in just three minutes on an Nvidia A100 GPU. It achieves accuracy on the order of $0.001$ to $0.01$ in inferred parameters for the first time. Additionally, Disk2Planet is capable of handling images with unknown noise levels, missing parts of arbitrary shapes, and various combinations of density and/or velocities, demonstrating its robustness in real observations.

\section{Method}\label{sec: method}

\begin{figure*}[tb!]
    \centering
    \includegraphics[width=0.9\textwidth]{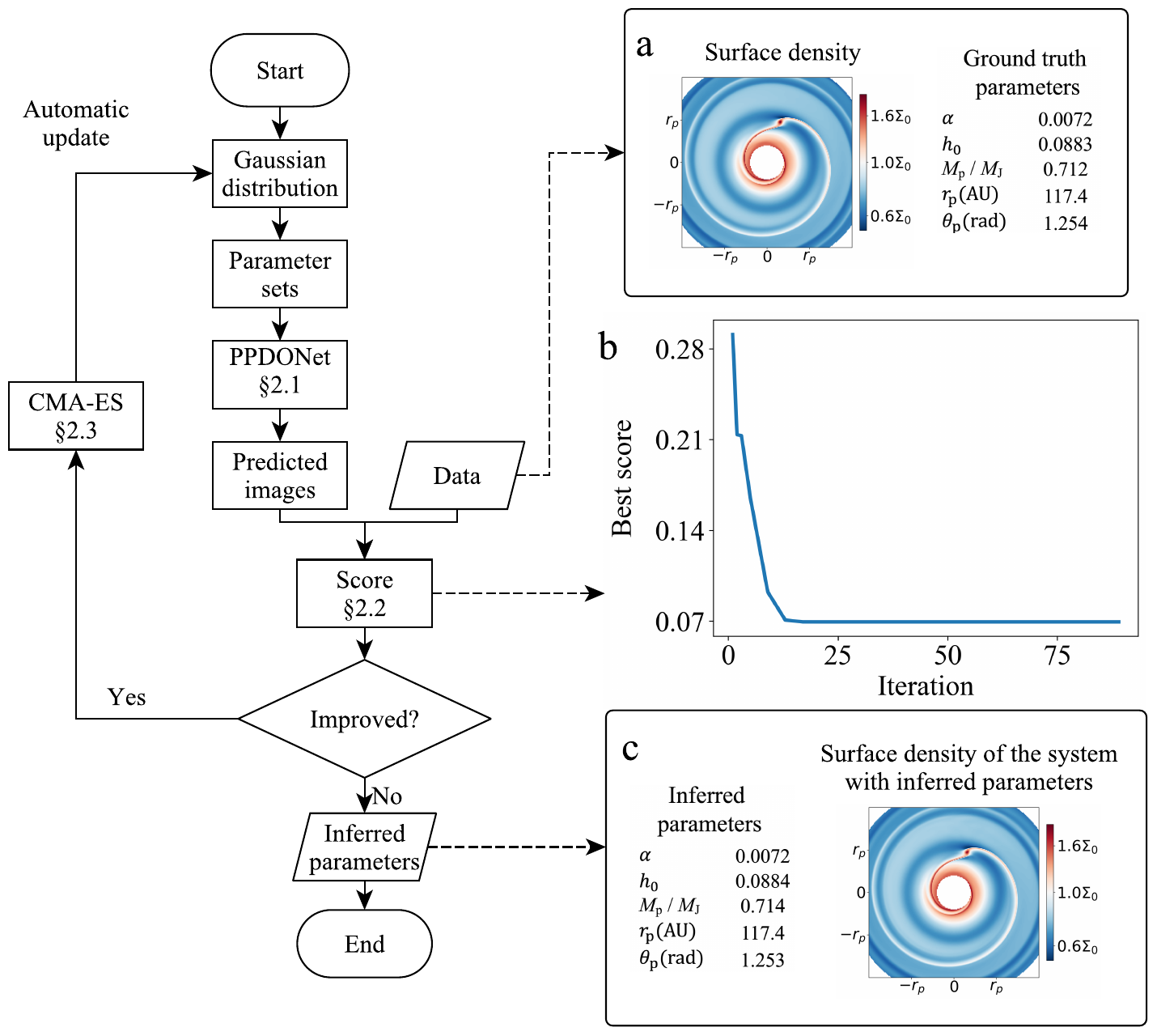}
    \caption{A flow chart illustrating the iterative parameter update procedure in Disk2Planet inverse problem solver. The solver iterates a 5D Gaussian distribution representing the most probable parameters (\(\alpha\), \(h_\mathrm{0}\), \(q\), \(r_\mathrm{p}\), \(\theta_\mathrm{p}\)) that result in the minimal difference between the model and the input data. At each iteration, the Gaussian distribution  is refined by testing 128 sets of parameters sampled from the current distribution. These parameters are processed by the ML-based solver PPDONet (\S\ref{sec: network}) to predict the corresponding disk maps. Differences between these maps and the data are quantified by the score (\S\ref{sec:compare}). The CMA--ES optimizer (\S\ref{sec: algorithm}) updates the Gaussian distribution to converge towards parameter sets with smaller scores. This process iterates until the distribution converges to the optimal parameters. On the right, we show an example of the inference process: (a) the input surface density map and its true parameters; (b) the evolution of the best score over iterations; (c) the inferred parameters and the corresponding surface density solution.
    \label{fig: flow chart}}
\end{figure*}

The inputs for our inverse problem solver consist of one or a combination of 2D maps of surface density and velocity distributions in a steady-state disk with a planet on a stable circular orbit. The solver is designed to find the optimal set of five parameters---Shakura--Sunyaev viscosity (\(\alpha\)), disk aspect ratio (\(h_\mathrm{0}\)), planet--star mass ratio (\(q\)), and planetary location (radius \(r_\mathrm{p}\) and azimuth \(\theta_\mathrm{p}\))---that results in the minimal differences between the predicted disk model and the input data. This is accomplished through an iterative process outlined in Fig.~\ref{fig: flow chart}.

In each iteration, we use a 5D Gaussian distribution to represent our current knowledge of (\(\alpha\), \(h_\mathrm{0}\), \(q\), \(r_\mathrm{p}\), \(\theta_\mathrm{p}\)). 
The mean of the Gaussian indicates the most probable parameter values.
Initially, we sample $4,096$ parameter sets uniformly in the parameter space of interest: $3\times10^{-4}\leq\alpha\leq0.1$, $0.05\leq h_\mathrm{0}\leq0.1$, $5\times10^{-5}\leq q\leq2\times10^{-3}$, and \(r_\mathrm{p}\) and \(\theta_\mathrm{p}\) within the input data image's boundaries (50--150 au for \(r_\mathrm{p}\) and $2\pi$ for \(\theta_\mathrm{p}\), or wherever the truncation due to cropping is; \S\ref{sec: results of cropped dataset}).
The mean of the Gaussian is initialized to the parameters with the lowest score (\S \ref{sec:compare}) among the $4,096$ parameter sets. The standard deviation of the Gaussian reflects the confidence in the estimate and is initially set to 0.01 \footnote{See the instructions at \url{https://CMA--ES.github.io/cmaes_sourcecode_page.html}.}.
Each iteration includes four steps: 
\begin{enumerate}
  \item Sample $128$ sets of parameters from the Gaussian distribution;
  \item Predict the disk model for each parameter set using the forward problem solver PPDONet (\S\ref{sec: network});
  \item Assign a score to each prediction to quantify how well it matches the input data (\S\ref{sec:compare});
  \item Update the Gaussian distribution's mean and standard deviation based on the scores using the Covariance Matrix Adaptation Evolution Strategy (CMA--ES; \S\ref{sec: algorithm}).
\end{enumerate}
The iterative process is repeated until the score in step (3) can no longer be minimized, indicating that the best possible set of parameters has been found.

Figure~\ref{fig: flow chart} illustrates how the tool works. The input data (panel (a)) contains a surface density map generated using the FARGO3D code \citep{masset2000fargo,benitez2016fargo3d} with its true parameters listed on the side. The evolution of the score is shown in panel (b). The inferred (optimal) parameters are shown in panel (c), where we achieve sub-percentage level accuracy for all five parameters. 
Panel (c) also shows the surface density map corresponding to the inferred parameters, obtained using PPDONet, enabling visual inspection.

\subsection{The PPDONet-based Forward Problem Solver}\label{sec: network}
To infer parameters automatically and accurately, our method tests many sets of parameters per iteration (e.g., 128 sets in our experiments) instead of a single set as in traditional approaches. This necessitates a fast forward problem solver, as traditional simulations are too slow to frequently test parameter sets. We employ PPDONet, the only public fast forward problem solver at the time of writing.

For each set of (\(\alpha\), \(h_\mathrm{0}\), \(q\)) sampled during iterations, PPDONet \citep{mao2023ppdonet} predicts surface density \(\Sigma\), radial velocity \(v_\mathrm{r}\), and azimuthal velocity \(v_\mathrm{\theta}\) maps, each computed in $0.01$ second on an Nvidia A100 chip. Line-of-sight velocity maps \(v_{\mathrm{LOS}} = v_r\cos(\theta) - v_{\theta}\sin(\theta)\) can be synthesized in post-processing. These maps are then rotated and stretched according to the planet location (\(r_\mathrm{p}\), \(\theta_\mathrm{p}\)) in the current sample. PPDONet is a publicly available machine learning tool\footnote{\url{https://github.com/smao-astro/PPDONet}} that efficiently maps scalar parameters from partial differential equations to their solutions. As demonstrated in \citet[Fig.~2]{mao2023ppdonet}, the predicted disk maps are visually indistinguishable from FARGO3D simulations within the parameter space of interest defined earlier.

\subsection{The Score for Data-Output Comparisons}\label{sec:compare}
The PPDONet-predicted disk map for each sampled parameter set is compared with the corresponding input data to generate a score quantifying their differences.
We opt for the L2 error metric over the structural similarity index measure (SSIM) \citep{wang2004image} due to its simplicity. In addition, our tests show that using L2 and SSIM as the evaluation metric yield similar results.
For velocities, the score is the relative \(L_2\) error:
\begin{equation}
L_2({\rm velocity}) = \| \text{pred} - \text{data} \|_2 / \| \text{data} \|_2,
\end{equation}
where \(\| \cdot \|_2\) denotes the root mean square over all pixel values. For surface density, the score is the logarithmic relative \(L_2\) error (e.g. \citealt{zhang2022pgnets})
\begin{equation}
L_2(\Sigma) = \| \log_{10}\text{pred} - \log_{10}\text{data} \|_2 / \| \log_{10}\text{data} \|_2.
\end{equation}

\begin{figure}[tb!]
    \centering
    \includegraphics[width=0.95\linewidth]{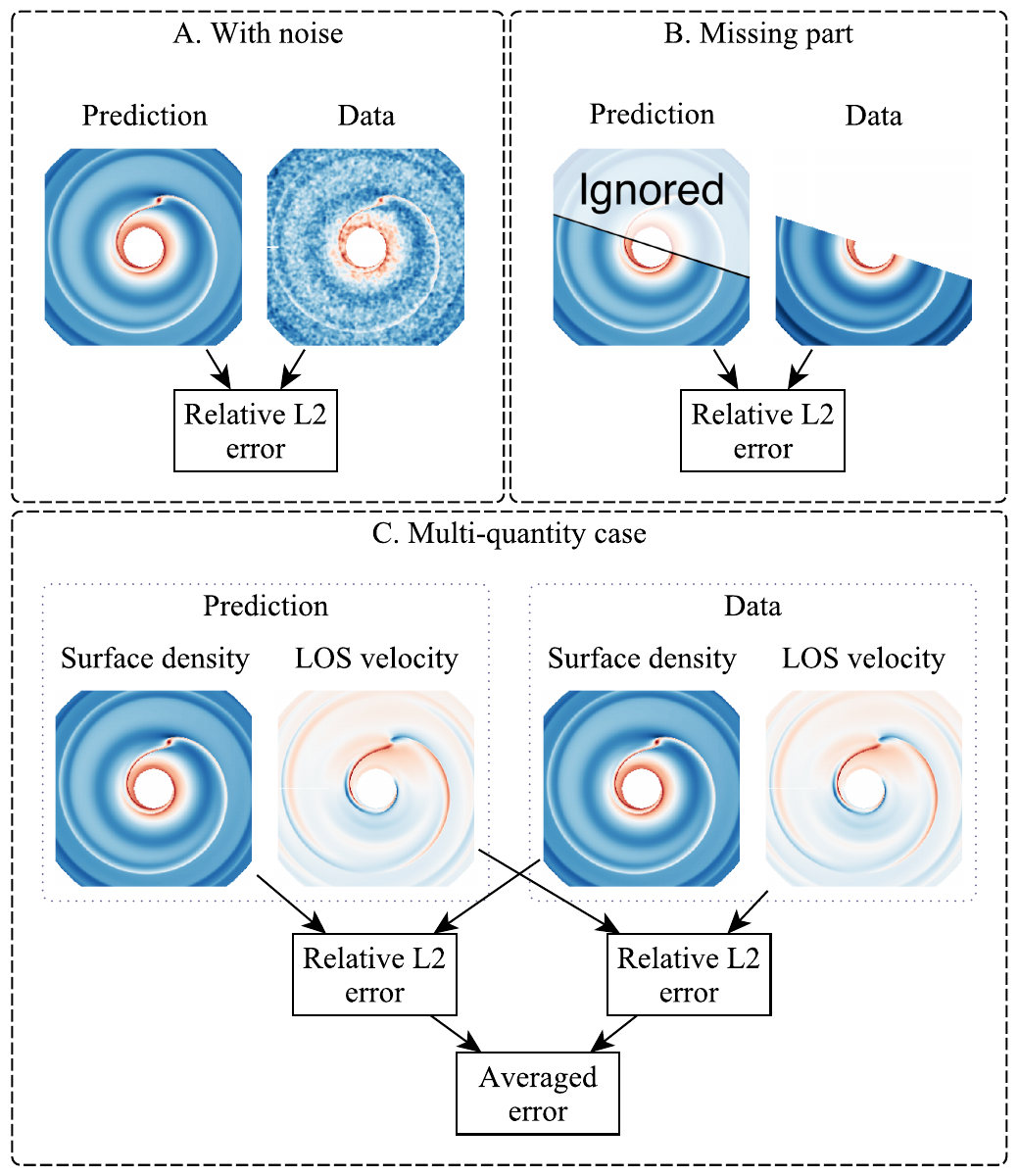}
    \caption{Score calculation for three cases of input data. Block A: For noisy input data, the score is calculated the same way as for noise-free data. Block B: For input data with missing areas, only the available pixels are used. Block C: For input data containing multiple quantities, the error for each quantity is calculated, and then the average of all errors is taken as the score.}\label{fig:score}
\end{figure}

The score is designed to handle the data complications illustrated in Fig.~\ref{fig:score}. For noisy data (block A), \(L_2\) is directly computed between the noisy data and the noise-free predicted model. For data with missing parts (block B), the error calculation is confined to valid pixels only. If the data includes multiple components, such as \(\Sigma\) and \(v_{\mathrm{LOS}}\) (block C), \(L_2\) for each component are averaged to produce a composite measure.
The score is not designed to account for effects caused by disk inclination, which is often known for individual systems \citep[e.g.,][Table 1]{huang2018disk}, enabling deprojection of the input data prior to parameter inferences when the observations probe the disk midplane (e.g., millimeter dust continuum emission).

\subsection{The Optimization Algorithm CMA--ES}\label{sec: algorithm}
We use an optimization algorithm to iterate parameter updates based on their corresponding scores. In optimization problems, there are two primary classes of algorithms: gradient-based and gradient-free. Gradient-based methods typically offer superior performance; however, they are unsuitable for our problem because the derivatives of scores with respect to parameters are unavailable. Consequently, we rely on a gradient-free technique: the Covariance Matrix Adaptation Evolution Strategy (CMA--ES) \citep{hansen1996adapting,hansen1997convergence,hansen2001completely,hansen2003reducing,hansen2006cma}. CMA--ES is renowned for its superior performance on benchmark problems \citep{hansen2010comparing,garcia2009study} where the correlation between parameters and scores is complex and not well understood analytically.

CMA--ES automatically updates the mean and covariance matrix of the Gaussian distribution to minimize the scores (\S\ref{sec:compare}) of the parameters sampled from the Gaussian distribution over iterations. Each update in CMA--ES follows an adaptive strategy similar to that found in evolutionary biology: The algorithm ranks parameter sets by their scores, selecting the lowest half as ``parents''. It then blends these parent sets through recombination and mutation to calculate the mean and covariance matrix for the new Gaussian distribution. This new Gaussian reflects the most successful parameter combinations at that moment, guiding the search towards promising directions. The updated Gaussian is then employed to generate ``children'' parameter sets for the next iteration.
This iterative process repeats until the score improvement plateaus, indicating that the optimal parameters have been effectively identified by the latest Gaussian distribution.

\section{Performance}\label{sec: experiments}
In this section, we evaluate the parameter inference accuracy of Disk2Planet. We begin with the baseline case in which the input data contains a complete noise-free map of surface density (\S\ref{sec:input-sigma}). We then test cases in which the input data contains multiple physical quantities (\S\ref{sec: results of multi dataset}), noises (\S\ref{sec: results of noisy dataset}), and missing parts (\S\ref{sec: results of cropped dataset}).

\subsection{The baseline case --- input data with surface density only}\label{sec:input-sigma}
\begin{figure}[tb]
    \centering
    \includegraphics[width=0.8\linewidth]{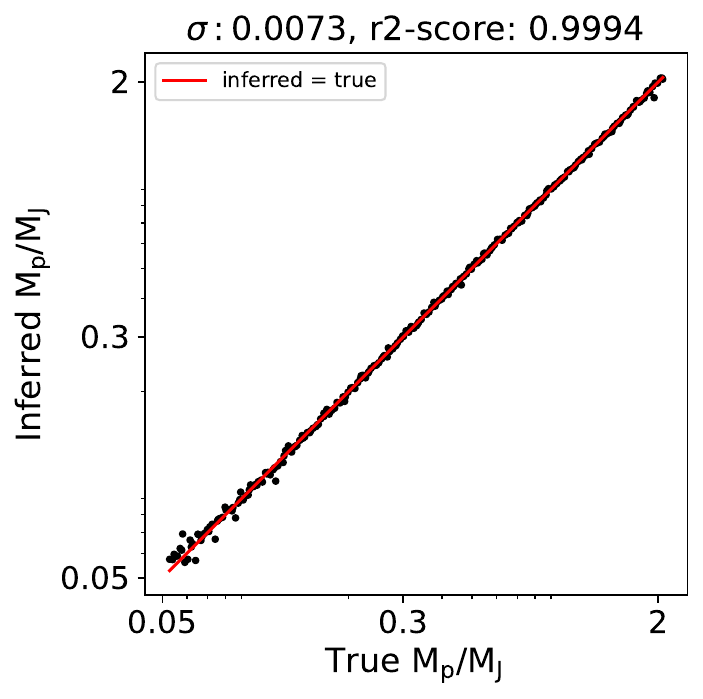}
    \caption{Ground truth and inferred planet masses in 256 tests using full noise-free maps of surface density as input. The tests achieve an $r2$-score (Eq.~\ref{eq:r2}) of 0.9994 and an inference uncertainty $\sigma$ of 0.0073, indicating near-perfect agreement (red solid line). See \S\ref{sec:input-sigma} for details.}\label{fig: q_trth_pred_line}
\end{figure}

Fig.~\ref{fig: q_trth_pred_line} shows the inferred versus ground truth planet masses for 256 tests uniformly sampled from the parameter space specified in \citet[][Table 1]{mao2023ppdonet}.Overall, the two agree well. The agreement can be quantitatively assessed using the \(r2\)-score: 
\begin{equation}
r2=1-SS(M_\mathrm{p}^{\mathrm{inferred}}-M_\mathrm{p}^{\mathrm{true}})/SS(M_\mathrm{p}^{\mathrm{true}}-\overline{M_\mathrm{p}^{\mathrm{true}}}),
\label{eq:r2}
\end{equation}
where $SS$ denotes the sum of square of all cases and $\overline{M_\mathrm{p}^{\mathrm{true}}}$ is the mean of all ground truth values. An \(r2\)-score close to one indicates near-perfect agreement. Our tool achieves an \(r2\)-score of \(0.9994\). For comparison, Bayesian network-based parameter inference achieves \(r2\)-scores of $0.82$  (\citealt{auddy2022using}, Fig.~3), and CNN-based parameter inference typically achieves \(r2\)-scores of $0.97$ to $0.98$ (\citealt{auddy2021dpnnet}, Fig.~4; \citealt{ruzza2024dbnets}, Fig.~5).

\begin{figure*}[tb]
    \centering
    \includegraphics[width=0.9\linewidth]{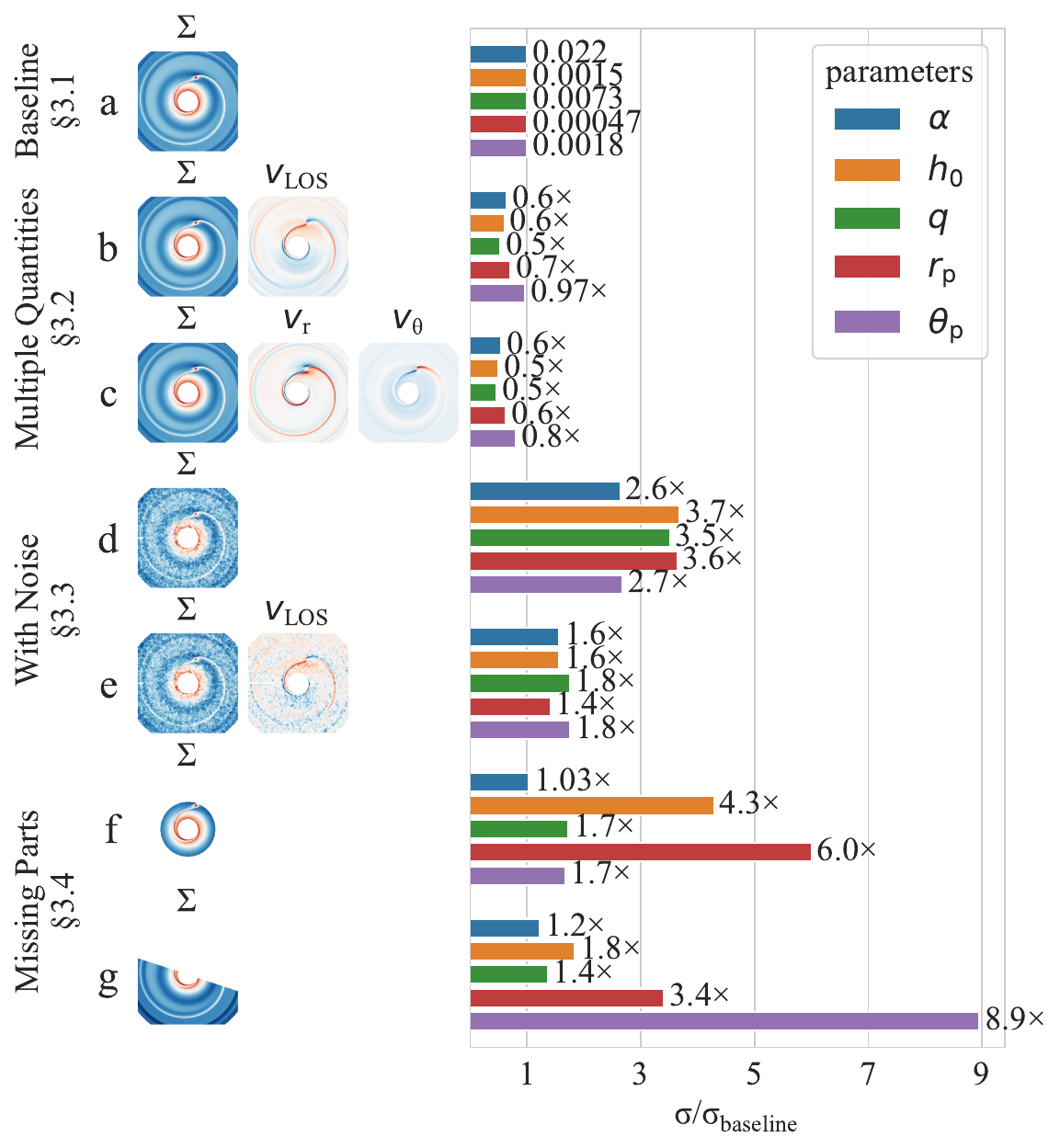}
    \caption{Representative examples of the input datasets, and the corresponding parameter inference uncertainties (\(\sigma\)). The input datasets contain images of, from top to bottom, \(\Sigma\) (a), \(\Sigma\)$+$\(v_\mathrm{LOS}\) (b), \(\Sigma\)$+$\(v_\mathrm{r}\)$+$\(v_\mathrm{\theta}\) (c), \(\Sigma\) with noise (d), \(\Sigma\)$+$\(v_\mathrm{LOS}\) with noise (e), radially cropped \(\Sigma\) (f), and azimuthally cropped \(\Sigma\) (g). 
    $256$ tests are run for each. The uncertainties are calculated as half the difference between the error distribution's $84^{\mathrm{th}}$ and $16^{\mathrm{th}}$ percentiles. The numbers on the bar plot represent the uncertainties for the baseline dataset (a), and how the uncertainties in the other datasets (b)--(g) compared to those in the baseline case.
    \label{fig: table with legend}}
\end{figure*}

Another metric for assessing the tool's performance is the error distribution of each parameter, as done by \citet{zhang2022pgnets} and \citet{ruzza2024dbnets}.
In each test, we quantify the error in each inferred parameter as:
\begin{subequations}
\begin{eqnarray}
    \mathrm{Err}(\alpha) & = & \log_{10}(\alpha^{\mathrm{inferred}}) - \log_{10}(\alpha^{\mathrm{true}}),\\
    \mathrm{Err}(h_\mathrm{0}) & = & (h_\mathrm{0}^{\mathrm{inferred}} - h_\mathrm{0}^{\mathrm{true}}) / h_\mathrm{0}^{\mathrm{true}},\\
    \mathrm{Err}(q) & = & \log_{10}(q^{\mathrm{inferred}}) - \log_{10}(q^{\mathrm{true}}),\\
    \mathrm{Err}(r_\mathrm{p}) & = & (r_\mathrm{p}^{\mathrm{inferred}} - r_\mathrm{p}^{\mathrm{true}}) / r_\mathrm{p}^{\mathrm{true}},\\
    \mathrm{Err}(\theta_\mathrm{p}) & = & \theta_\mathrm{p}^{\mathrm{inferred}} - \theta_\mathrm{p}^{\mathrm{true}}.
\end{eqnarray}
\label{eq:errors}
\end{subequations}
The errors for $\alpha$ and $q$ are calculated on a logarithmic scale due to their wide range across several orders of magnitude \citep[Eq. 13]{ruzza2024dbnets}. The errors for $h_\mathrm{0}$ and $r_\mathrm{p}$ are normalized. We then define the inference uncertainty $\sigma$ as half the difference between the error distribution's $84^{\mathrm{th}}$ and $16^{\mathrm{th}}$ percentiles, and list them in the first column in Fig.~\ref{fig: table with legend}: $\sigma(\alpha)=0.022$, $\sigma(h_\mathrm{0})=0.0015$, $\sigma(q)=0.0073$, $\sigma(r_\mathrm{p})=0.00047$, $\sigma(\theta_\mathrm{p})=0.0018$. In comparison, CNN-based parameter inference achieves $\sigma(q)=0.16$ \citep{zhang2022pgnets} and $0.13$ \citep{ruzza2024dbnets}, and $\sigma(\alpha)=0.23$ \citep{zhang2022pgnets}.

\begin{figure*}[tb]
    \centering
    \epsscale{1}
    \plottwo{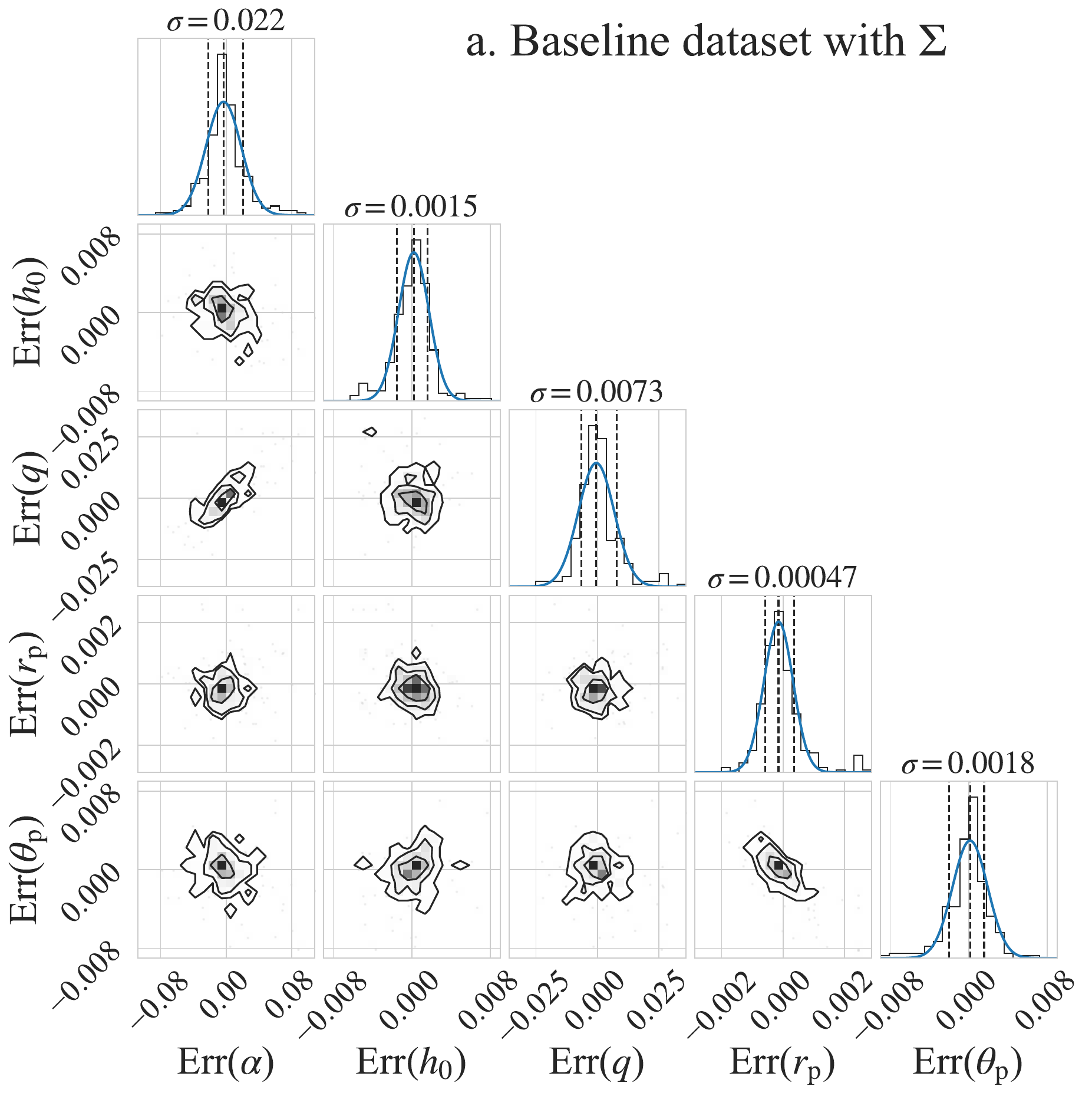}{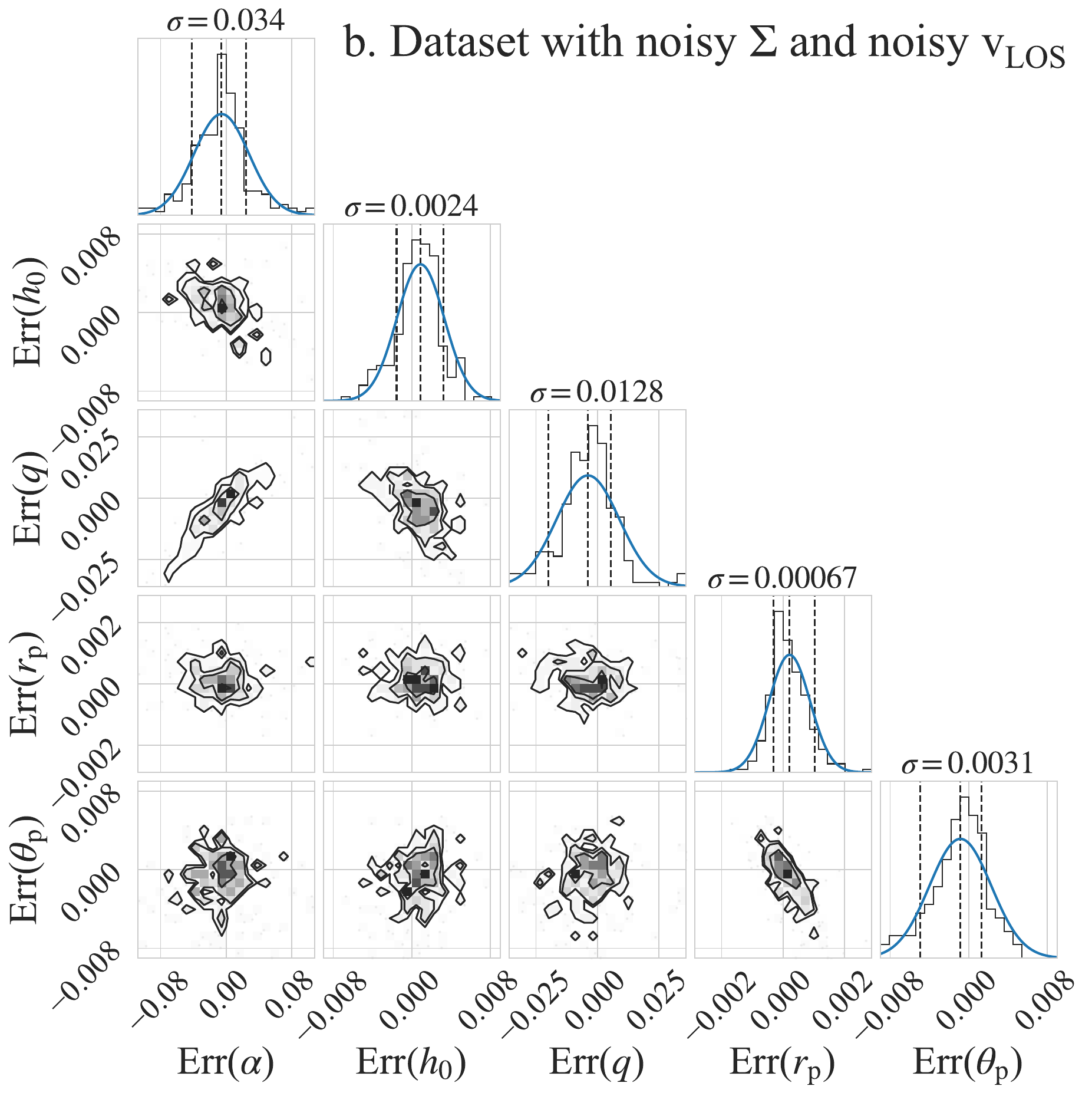}
    \caption{Corner plots showing the distribution of errors in the inferred parameters (Eq.~\ref{eq:errors}). Corresponding panels on the two sides are on the same scales. The histograms on the diagonal show the distributions of individual parameters, while the off-diagonal panels highlight pairwise correlations. See \S\ref{sec:input-sigma} for details.
    \label{fig: corner plot}}
\end{figure*}

We examine the error distributions and possible correlations in the multi-dimensional parameter space in Fig.~\ref{fig: corner plot}a. The histograms on the diagonal show the distributions of individual parameters while the off-diagonal panels highlight pairwise correlations. Notably, all distributions are centered, indicating little systematic bias in the inferred parameters. Additionally, only two pairwise correlations (out of a total of $C(5,2)=10$) are not isotropic: $\mathrm{Err}(\alpha)$ {\it vs} $\mathrm{Err}(q)$ exhibits a positive correlation, echoing \citet[Fig.~5]{zhang2022pgnets}, while $\mathrm{Err}(r_\mathrm{p})$ is negatively correlated with $\mathrm{Err} (\theta_\mathrm{p})$.

\subsection{Input data with multiple quantities}\label{sec: results of multi dataset}
Disk2Planet can incorporate surface density \(\Sigma\), radial velocity \(v_\mathrm{r}\), and azimuthal velocity \(v_\mathrm{\theta}\) in any combination. For example, we can supply two maps, \(\Sigma\) $+$ \(v_{\mathrm{LOS}}\), as illustrated in Fig.~\ref{fig: table with legend} (case (b)). This scenario mimics real situations where gas line emission observations (e.g., \citealt{zhang2021molecules}) provide these quantities. As the amount of information in the input data increases, the parameter inference accuracy improves, resulting in smaller errors. Compared to the baseline case (a), adding \(v_{\mathrm{LOS}}\) reduces uncertainties for the five inferred parameters by $3$\% to $47$\% in 256 tests. Additionally, supplying three maps, \(\Sigma\) $+$ \(v_\mathrm{r}\) $+$ \(v_\mathrm{\theta}\), as input data (case c) further reduces uncertainties since \(v_\mathrm{r}\) $+$ \(v_\mathrm{\theta}\) provide more information than \(v_{\mathrm{LOS}}\).

\subsection{Input data with noise}\label{sec: results of noisy dataset}
Observational data often contain noise. Our inverse problem solver can handle noisy input data, as shown in cases (d) and (e) in Fig.~\ref{fig: table with legend}. They are created by introducing random uniform noises in \(\Sigma\) (case (d)) and $\Sigma+v_{\mathrm{LOS}}$ (case (e)). The noises
have a spatial scale of 2.5~AU (\citealt[][\S2.3;]{zhang2018disk} equivalent to 0.018 arcsec at 140 pc), added to the logarithmic pixel values with a standard deviation $0.6\times$ the average of the quantity. The resulting uncertainties for the five parameters in both cases are $2-4\times$ those in the corresponding noise-free inputs (cases (a) and (b)). The associated multi-dimensional error distributions are shown in Fig.~\ref{fig: corner plot}b, with trends similar to those in the baseline case (Fig.~\ref{fig: corner plot}a; \S\ref{sec:input-sigma}).

\subsection{Input data with missing parts}\label{sec: results of cropped dataset}
In observations, signals in certain regions may be missing or masked for various reasons, such as foreground extinction \citep{bruderer2014gas, tsukagoshi2019flared}. Figure~\ref{fig: table with legend} illustrates two such cases: the \(\Sigma\) map is cropped radially beyond \(1.1 r_\mathrm{p}\) (f) and azimuthally in half (g; the planet is in the cropped region). In both scenarios, the tool performs parameter inference in all 256 tests, albeit with uncertainties $1-9\times$ those in the baseline case with a full \(\Sigma\) map (a). This is expected, as cropping reduces the amount of information in the input image. In the radial cropping case, $\sigma(h_\mathrm{0})$ and $\sigma(r_\mathrm{p})$ have the largest increases ($4.3\times$ and $6.0\times$, respectively), although their absolute values remain low ($0.007$ and $0.003$, respectively). In the azimuthal cropping case, a 9--fold increase in $\sigma(\theta_\mathrm{p})$ stands out among the five uncertainties, attributed to the loss of substructures near the planet, crucial for its accurate localization. The absolute value of $\sigma(\theta_\mathrm{p})$, \(0.016\) (1$^\circ$, or 1.6 AU for a planet at 100 AU), remains smaller than the typical resolution achieved in today's high-resolution imaging observations of disks \citep[e.g.,][]{benisty2022optical}.

\section{Advantages over existing inverse problem solvers}\label{sec:advantages}
Disk2Planet is robust against noisy data (\ref{sec: results of noisy dataset}). This is because our optimization focuses on large-scale disk morphology (e.g., spiral arms, gap), more than noises (usually small-scale). To demonstrate this, imagine a small error in the inferred disk scale height. This will cause noticeable differences in the spiral arm pitch angle \citep{fung2015inferring}, which is much more than the influence of the noise (\S\ref{sec:compare}). Thus Disk2Planet is generally noise-robust.
In contrast, CNN-based methods \citep{auddy2021dpnnet,zhang2022pgnets,ruzza2024dbnets} rely on {\it local} spatial patterns, thus are more prone to noises. They result in significantly larger errors when the signal-to-noise ratio is below ten (e.g., a planet mass error of $0.68$ dex versus $0.13$ dex; \citealt[\S 5.3]{ruzza2024dbnets}).

Our inverse problem solver can process incomplete images with missing parts in arbitrary shapes in a straightforward manner. As illustrated in Fig.~\ref{fig:score} (Block B), the missing parts are ignored and do not contribute to the parameter inference. In contrast, CNN-based methods require completed images, as every pixel contributes to the convolution, which is key to CNNs in extracting information. In theory, CNNs can also be trained with datasets that have the same missing parts as the real observations. However, this is impractical in real-life applications, because it is often difficult to predict which parts of the image will be missing in real observations, and the missing parts can vary between observations.

Our inverse problem solver works with input images of any spatial resolution since the entire procedure is resolution-independent. In contrast, CNN-based solvers are often trained on datasets with fixed resolutions. To minimize training costs, low resolutions are typically used, such as \(64 \times 64\) in \citet{zhang2022pgnets} or \(128 \times 128\) in \citet{ruzza2024dbnets}, which can degrade detailed disk substructures like spiral arms.  Additionally, due to limitations in the training sets, CNN-based tools can only operate properly when the input images have the same resolution as the training sets; if not, regridding is needed, which may result in a loss of information.

Disk2Planet locates planets by matching their induced substructures with those in the data. This contrasts with CNN-based approaches \citep[e.g.,][]{ruzza2024dbnets}, which requires users to provide prior knowledge of the planet's location. Augmenting the training dataset in PPDONet to accommodate different planet locations is not needed neither, in contrast to CNN-based approach (e.g. \citealt{zhang2022pgnets}). 

Disk2Planet has an advantage over the more traditional CNN-based methods when dealing with input images with parameters outside the specified range. Our method leverages the fast forward solver, PPDONet, which generates a modelled image (Fig.~\ref{fig: flow chart}, panel (c)) based on the inferred parameters at no additional computational cost. This feature enables users to visually compare the modelled image with the input image. 
When the input image parameters fall outside the specified range, the inferred parameters are constrained within the predefined parameter space by our tool, thus lead the modelled image differs from the input image. Consequently, visual comparison provides an intuitive means of verifying results and identifying potential issues with input data.
In comparison, CNN-based approach doesn't offer a straightforward approach to visually verify the results.

\section{Conclusions and future perspectives}\label{sec: conclusion}
We have developed a fast and fully automated inverse problem solver, Disk2Planet, capable of inferring five parameters --- the Shakura--Sunyaev viscosity (\(\alpha\)), disk aspect ratio (\(h_\mathrm{0}\)), planet--star mass ratio (\(q\)), and the planet's radius and azimuth --- from 2D steady-state disk-planet systems. Inferring these parameters from surface density in one system takes three GPU minutes on an Nvidia A100 machine. Our solver requires minimal human intervention and is user-friendly for newcomers without prior experience in numerical simulations of disk--planet interactions.

The architecture of the solver and an example application are shown in Fig.~\ref{fig: flow chart}.
The solver is built on the machine learning tool PPDONet, which predicts the disk maps created by a planet \citep{mao2023ppdonet}.
The solver employs the evolutionary optimization algorithm CMA--ES \citep{hansen1996adapting,hansen1997convergence,hansen2001completely,hansen2003reducing,hansen2006cma,hansen2019pycma}, which samples parameters from a 5D Gaussian distribution in each iteration and refines the distribution by minimizing the difference between the model and the input data.

The inputs to the solver consist of 2D maps of gas surface density and velocities in a disk. The solver can handle data with unknown noise levels, incomplete data with missing parts, and user-defined combinations of quantities (Fig.~\ref{fig: table with legend}). It achieves percentage-level or smaller errors in the inferred parameters in the tests performed in this work (Fig.~\ref{fig: q_trth_pred_line}), representing at least an order of magnitude improvement over previous tools based on Convolutional Neural Networks \citep[e.g.,][]{auddy2021dpnnet,zhang2022pgnets,ruzza2024dbnets}. As expected, the errors decrease as more information is provided by the input data.

The current version of Disk2Planet may be improved in certain ways. First, it is designed to work with 2D maps of gas surface density and velocities derived from observations \citep[e.g.,][]{zhang2021molecules}, rather than directly with observational data. To extend its functionality to disk observations directly, an updated PPDONet capable of directly predicting synthetic images is required. Training such a network would necessitate a dataset generated from gas+dust simulations and synthetic observations, and sampling additional parameters such as dust properties and planet masses and orbits for multi-planet systems. 
Once such a dataset is available, future work can readily adapt our approach to update the existing inverse problem solver. These modifications would require minimal changes to the existing PPDONet architecture and our inverse problem solver framework.
Second, unlike \citet{auddy2022using} and \citet{ruzza2024dbnets}, it does not produce uncertainties on the inferred parameters. However, the forward problem solver employed, PPDONet, provides the predicted disk maps corresponding to the inferred parameters (Fig.~\ref{fig: flow chart}). These disk maps can be visually compared to data, offering users a straightforward way to evaluate the quality of the inferences.

The data and source code for this work will be available at \url{https://github.com/smao-astro/Disk2Planet} upon publication.

\begin{acknowledgments}
We are grateful to an anonymous referee for constructive suggestions that improved our paper.
We thank Jaehan Bae, Xuening Bai, Pablo Ben{\'\i}tez-Llambay, Shengze Cai, Miles Cranmer, Bin Dong, Scott Field, Jeffrey Fung, Xiaotian Gao, Jiequn Han, Pinghui Huang, Pengzhan Jin, Xiaowei Jin, Hui Li, Tie-Yan Liu, Zhiping Mao, Chris Ormel, Wenlei Shi, Karun Thanjavur, Yiwei Wang, Yinhao Wu, Zhenghao Xu, Minhao Zhang, Wei Zhu for help and useful discussions in the project.

S.M. and R.D. are supported by the Natural Sciences and Engineering Research Council of Canada (NSERC) and the Alfred P. Sloan Foundation. S.M. and R.D. acknowledge the support of the Government of Canada's New Frontiers in Research Fund (NFRF), [NFRFE-2022-00159].
This research was enabled in part by support provided by the Digital Research Alliance of Canada \url{alliance.can.ca}.
\end{acknowledgments}

\bibliography{main}{}

\begin{thebibliography}{}
\expandafter\ifx\csname natexlab\endcsname\relax\def\natexlab#1{#1}\fi
\providecommand{\url}[1]{\href{#1}{#1}}
\providecommand{\dodoi}[1]{doi:~\href{http://doi.org/#1}{\nolinkurl{#1}}}
\providecommand{\doeprint}[1]{\href{http://ascl.net/#1}{\nolinkurl{http://ascl.net/#1}}}
\providecommand{\doarXiv}[1]{\href{https://arxiv.org/abs/#1}{\nolinkurl{https://arxiv.org/abs/#1}}}

\bibitem[{Andrews(2020)}]{andrews2020observations}
Andrews, S.~M. 2020, Annual Review of Astronomy and Astrophysics, 58, 483

\bibitem[{Auddy {et~al.}(2022)Auddy, Dey, Lin, Carrera, \& Simon}]{auddy2022using}
Auddy, S., Dey, R., Lin, M.-K., Carrera, D., \& Simon, J.~B. 2022, The Astrophysical Journal, 936, 93

\bibitem[{Auddy {et~al.}(2021)Auddy, Dey, Lin, \& Hall}]{auddy2021dpnnet}
Auddy, S., Dey, R., Lin, M.-K., \& Hall, C. 2021, The Astrophysical Journal, 920, 3

\bibitem[{Auddy \& Lin(2020)}]{auddy2020machine}
Auddy, S., \& Lin, M.-K. 2020, The Astrophysical Journal, 900, 62

\bibitem[{Bae \& Zhu(2018)}]{bae2018planet}
Bae, J., \& Zhu, Z. 2018, The Astrophysical Journal, 859, 118

\bibitem[{Benisty {et~al.}(2022)Benisty, Dominik, Follette, Garufi, Ginski, Hashimoto, Keppler, Kley, \& Monnier}]{benisty2022optical}
Benisty, M., Dominik, C., Follette, K., {et~al.} 2022, arXiv preprint arXiv:2203.09991

\bibitem[{Ben{\'\i}tez-Llambay \& Masset(2016)}]{benitez2016fargo3d}
Ben{\'\i}tez-Llambay, P., \& Masset, F.~S. 2016, The Astrophysical Journal Supplement Series, 223, 11

\bibitem[{Bruderer {et~al.}(2014)Bruderer, van~der Marel, Van~Dishoeck, \& van Kempen}]{bruderer2014gas}
Bruderer, S., van~der Marel, N., Van~Dishoeck, E.~F., \& van Kempen, T.~A. 2014, Astronomy \& Astrophysics/Astronomie et Astrophysique, 562

\bibitem[{Christiaens {et~al.}(2019)Christiaens, Casassus, Absil, Cantalloube, Gomez~Gonzalez, Girard, Ram{\'\i}rez, Pairet, Salinas, Price, {et~al.}}]{christiaens2019separating}
Christiaens, V., Casassus, S., Absil, O., {et~al.} 2019, Monthly Notices of the Royal Astronomical Society, 486, 5819

\bibitem[{Cilibrasi {et~al.}(2023)Cilibrasi, Flock, \& Szul{\'a}gyi}]{cilibrasi2023meridional}
Cilibrasi, M., Flock, M., \& Szul{\'a}gyi, J. 2023, Monthly Notices of the Royal Astronomical Society, 523, 2039

\bibitem[{Currie {et~al.}(2022)Currie, Lawson, Schneider, Lyra, Wisniewski, Grady, Guyon, Tamura, Kotani, Kawahara, {et~al.}}]{currie2022images}
Currie, T., Lawson, K., Schneider, G., {et~al.} 2022, Nature Astronomy, 6, 751

\bibitem[{Dipierro \& Laibe(2017)}]{dipierro2017opening}
Dipierro, G., \& Laibe, G. 2017, Monthly Notices of the Royal Astronomical Society, 469, 1932

\bibitem[{Dipierro {et~al.}(2015)Dipierro, Price, Laibe, Hirsh, Cerioli, \& Lodato}]{dipierro2015planet}
Dipierro, G., Price, D., Laibe, G., {et~al.} 2015, Monthly Notices of the Royal Astronomical Society: Letters, 453, L73

\bibitem[{Dong \& Fung(2017)}]{dong2017bright}
Dong, R., \& Fung, J. 2017, The Astrophysical Journal, 835, 38

\bibitem[{Dong {et~al.}(2018)Dong, Li, Chiang, \& Li}]{dong2018multiple}
Dong, R., Li, S., Chiang, E., \& Li, H. 2018, The Astrophysical Journal, 866, 110

\bibitem[{Dong {et~al.}(2015{\natexlab{a}})Dong, Zhu, Rafikov, \& Stone}]{dong2015observational}
Dong, R., Zhu, Z., Rafikov, R.~R., \& Stone, J.~M. 2015{\natexlab{a}}, The Astrophysical Journal Letters, 809, L5

\bibitem[{Dong {et~al.}(2015{\natexlab{b}})Dong, Zhu, \& Whitney}]{dong2015observational2}
Dong, R., Zhu, Z., \& Whitney, B. 2015{\natexlab{b}}, The Astrophysical Journal, 809, 93

\bibitem[{Fung \& Dong(2015)}]{fung2015inferring}
Fung, J., \& Dong, R. 2015, The Astrophysical Journal Letters, 815, L21

\bibitem[{Fung {et~al.}(2014)Fung, Shi, \& Chiang}]{fung2014empty}
Fung, J., Shi, J.-M., \& Chiang, E. 2014, The Astrophysical Journal, 782, 88

\bibitem[{Garc{\'\i}a {et~al.}(2009)Garc{\'\i}a, Molina, Lozano, \& Herrera}]{garcia2009study}
Garc{\'\i}a, S., Molina, D., Lozano, M., \& Herrera, F. 2009, Journal of Heuristics, 15, 617

\bibitem[{Haffert {et~al.}(2019)Haffert, Bohn, De~Boer, Snellen, Brinchmann, Girard, Keller, \& Bacon}]{haffert2019two}
Haffert, S., Bohn, A., De~Boer, J., {et~al.} 2019, Nature Astronomy, 3, 749

\bibitem[{Hansen(2006)}]{hansen2006cma}
Hansen, N. 2006, Towards a new evolutionary computation: Advances in the estimation of distribution algorithms, 75

\bibitem[{Hansen {et~al.}(2019)Hansen, Akimoto, \& Baudis}]{hansen2019pycma}
Hansen, N., Akimoto, Y., \& Baudis, P. 2019, {CMA-ES/pycma} on {G}ithub, Zenodo, DOI:10.5281/zenodo.2559634, \dodoi{10.5281/zenodo.2559634}

\bibitem[{Hansen {et~al.}(2010)Hansen, Auger, Ros, Finck, \& Po{\v{s}}{\'\i}k}]{hansen2010comparing}
Hansen, N., Auger, A., Ros, R., Finck, S., \& Po{\v{s}}{\'\i}k, P. 2010, in Proceedings of the 12th annual conference companion on Genetic and evolutionary computation, 1689--1696

\bibitem[{Hansen {et~al.}(2003)Hansen, M{\"u}ller, \& Koumoutsakos}]{hansen2003reducing}
Hansen, N., M{\"u}ller, S.~D., \& Koumoutsakos, P. 2003, Evolutionary computation, 11, 1

\bibitem[{Hansen \& Ostermeier(1996)}]{hansen1996adapting}
Hansen, N., \& Ostermeier, A. 1996, in Proceedings of IEEE international conference on evolutionary computation, IEEE, 312--317

\bibitem[{Hansen \& Ostermeier(1997)}]{hansen1997convergence}
Hansen, N., \& Ostermeier, A. 1997, Eufit, 97, 650

\bibitem[{Hansen \& Ostermeier(2001)}]{hansen2001completely}
---. 2001, Evolutionary computation, 9, 159

\bibitem[{Hashimoto {et~al.}(2020)Hashimoto, Aoyama, Konishi, Uyama, Takasao, Ikoma, \& Tanigawa}]{hashimoto2020accretion}
Hashimoto, J., Aoyama, Y., Konishi, M., {et~al.} 2020, The Astronomical Journal, 159, 222

\bibitem[{Huang {et~al.}(2018)Huang, Andrews, Dullemond, Isella, P{\'e}rez, Guzm{\'a}n, {\"O}berg, Zhu, Zhang, Bai, {et~al.}}]{huang2018disk}
Huang, J., Andrews, S.~M., Dullemond, C.~P., {et~al.} 2018, The Astrophysical Journal Letters, 869, L42

\bibitem[{Izquierdo {et~al.}(2021)Izquierdo, Testi, Facchini, Rosotti, \& van Dishoeck}]{izquierdo2021disc}
Izquierdo, A.~F., Testi, L., Facchini, S., Rosotti, G.~P., \& van Dishoeck, E.~F. 2021, arXiv preprint arXiv:2104.09596

\bibitem[{Jin {et~al.}(2016)Jin, Li, Isella, Li, \& Ji}]{jin2016modeling}
Jin, S., Li, S., Isella, A., Li, H., \& Ji, J. 2016, The Astrophysical Journal, 818, 76

\bibitem[{Keppler {et~al.}(2018)Keppler, Benisty, M{\"u}ller, Henning, Van~Boekel, Cantalloube, Ginski, Van~Holstein, Maire, Pohl, {et~al.}}]{keppler2018discovery}
Keppler, M., Benisty, M., M{\"u}ller, A., {et~al.} 2018, Astronomy \& Astrophysics, 617, A44

\bibitem[{Li {et~al.}(2021)Li, Liu, Yang, Peng, \& Zhou}]{li2021survey}
Li, Z., Liu, F., Yang, W., Peng, S., \& Zhou, J. 2021, IEEE transactions on neural networks and learning systems

\bibitem[{Mao {et~al.}(2023)Mao, Dong, Lu, Yi, Wang, \& Perdikaris}]{mao2023ppdonet}
Mao, S., Dong, R., Lu, L., {et~al.} 2023, The Astrophysical Journal Letters, 950, L12

\bibitem[{Masset(2000)}]{masset2000fargo}
Masset, F. 2000, Astronomy and Astrophysics Supplement Series, 141, 165

\bibitem[{M{\"u}ller {et~al.}(2018)M{\"u}ller, Keppler, Henning, Samland, Chauvin, Beust, Maire, Molaverdikhani, van Boekel, Benisty, {et~al.}}]{muller2018orbital}
M{\"u}ller, A., Keppler, M., Henning, T., {et~al.} 2018, Astronomy \& Astrophysics, 617, L2

\bibitem[{O'Shea \& Nash(2015)}]{o2015introduction}
O'Shea, K., \& Nash, R. 2015, arXiv preprint arXiv:1511.08458

\bibitem[{Paardekooper {et~al.}(2022)Paardekooper, Dong, Duffell, Fung, Masset, Ogilvie, \& Tanaka}]{paardekooper2022planet}
Paardekooper, S.-J., Dong, R., Duffell, P., {et~al.} 2022, arXiv preprint arXiv:2203.09595

\bibitem[{Paardekooper \& Mellema(2006)}]{paardekooper2006dust}
Paardekooper, S.-J., \& Mellema, G. 2006, arXiv preprint astro-ph/0603132

\bibitem[{Pinte {et~al.}(2018)Pinte, Price, M{\'e}nard, Duch{\^e}ne, Dent, Hill, de~Gregorio-Monsalvo, Hales, \& Mentiplay}]{pinte2018kinematic}
Pinte, C., Price, D., M{\'e}nard, F., {et~al.} 2018, The Astrophysical Journal Letters, 860, L13

\bibitem[{Pinte {et~al.}(2019)Pinte, van Der~Plas, M{\'e}nard, Price, Christiaens, Hill, Mentiplay, Ginski, Choquet, Boehler, {et~al.}}]{pinte2019kinematic}
Pinte, C., van Der~Plas, G., M{\'e}nard, F., {et~al.} 2019, Nature Astronomy, 3, 1109

\bibitem[{Rabago \& Zhu(2021)}]{rabago2021constraining}
Rabago, I., \& Zhu, Z. 2021, Monthly Notices of the Royal Astronomical Society, 502, 5325

\bibitem[{Rosotti {et~al.}(2016)Rosotti, Juhasz, Booth, \& Clarke}]{rosotti2016minimum}
Rosotti, G.~P., Juhasz, A., Booth, R.~A., \& Clarke, C.~J. 2016, Monthly Notices of the Royal Astronomical Society, 459, 2790

\bibitem[{Ruzza {et~al.}(2024)Ruzza, Lodato, \& Rosotti}]{ruzza2024dbnets}
Ruzza, A., Lodato, G., \& Rosotti, G.~P. 2024, arXiv preprint arXiv:2402.12448

\bibitem[{Teague {et~al.}(2018)Teague, Bae, Bergin, Birnstiel, \& Foreman-Mackey}]{teague2018kinematical}
Teague, R., Bae, J., Bergin, E.~A., Birnstiel, T., \& Foreman-Mackey, D. 2018, The Astrophysical Journal Letters, 860, L12

\bibitem[{Terry {et~al.}(2022)Terry, Hall, Abreau, \& Gleyzer}]{terry2022locating}
Terry, J., Hall, C., Abreau, S., \& Gleyzer, S. 2022, The Astrophysical Journal, 941, 192

\bibitem[{Tsukagoshi {et~al.}(2019)Tsukagoshi, Momose, Kitamura, Saito, Kawabe, Andrews, Wilner, Kudo, Hashimoto, Ohashi, {et~al.}}]{tsukagoshi2019flared}
Tsukagoshi, T., Momose, M., Kitamura, Y., {et~al.} 2019, The Astrophysical Journal, 871, 5

\bibitem[{Wagner {et~al.}(2018)Wagner, Follette, Close, Apai, Gibbs, Keppler, M{\"u}ller, Henning, Kasper, Wu, {et~al.}}]{wagner2018magellan}
Wagner, K., Follette, K.~B., Close, L.~M., {et~al.} 2018, The Astrophysical Journal Letters, 863, L8

\bibitem[{Wagner {et~al.}(2023)Wagner, Stone, Skemer, Ertel, Dong, Apai, Spalding, Leisenring, Sitko, Kratter, {et~al.}}]{wagner2023direct}
Wagner, K., Stone, J., Skemer, A., {et~al.} 2023, Nature Astronomy, 7, 1208

\bibitem[{Wang {et~al.}(2020)Wang, Ginzburg, Ren, Wallack, Gao, Mawet, Bond, Cetre, Wizinowich, De~Rosa, {et~al.}}]{wang2020keck}
Wang, J.~J., Ginzburg, S., Ren, B., {et~al.} 2020, The Astronomical Journal, 159, 263

\bibitem[{Wang {et~al.}(2004)Wang, Bovik, Sheikh, \& Simoncelli}]{wang2004image}
Wang, Z., Bovik, A.~C., Sheikh, H.~R., \& Simoncelli, E.~P. 2004, IEEE transactions on image processing, 13, 600

\bibitem[{Zhang {et~al.}(2021)Zhang, Booth, Law, Bosman, Schwarz, Bergin, {\"O}berg, Andrews, Guzm{\'a}n, Walsh, {et~al.}}]{zhang2021molecules}
Zhang, K., Booth, A.~S., Law, C.~J., {et~al.} 2021, The Astrophysical Journal Supplement Series, 257, 5

\bibitem[{Zhang {et~al.}(2022)Zhang, Zhu, \& Kang}]{zhang2022pgnets}
Zhang, S., Zhu, Z., \& Kang, M. 2022, Monthly Notices of the Royal Astronomical Society, 510, 4473

\bibitem[{Zhang {et~al.}(2018)Zhang, Zhu, Huang, Guzm{\'a}n, Andrews, Birnstiel, Dullemond, Carpenter, Isella, P{\'e}rez, {et~al.}}]{zhang2018disk}
Zhang, S., Zhu, Z., Huang, J., {et~al.} 2018, The Astrophysical Journal Letters, 869, L47

\bibitem[{Zhou {et~al.}(2022)Zhou, Sanghi, Bowler, Wu, Close, Long, Ward-Duong, Zhu, Kraus, Follette, {et~al.}}]{zhou2022hst}
Zhou, Y., Sanghi, A., Bowler, B.~P., {et~al.} 2022, The Astrophysical Journal Letters, 934, L13

\bibitem[{Zhu {et~al.}(2014)Zhu, Stone, Rafikov, \& Bai}]{zhu2014particle}
Zhu, Z., Stone, J.~M., Rafikov, R.~R., \& Bai, X.-n. 2014, The Astrophysical Journal, 785, 122

\end{thebibliography}
\bibliographystyle{aasjournal}



\end{document}